\documentclass[]{spie}  

\usepackage{bm}
\usepackage{amsmath,amsfonts,amssymb}
\usepackage{graphicx}
\usepackage{tikz}
\usepackage{subcaption}
\usepackage[percent]{overpic}
\usepackage[colorlinks=true, allcolors=blue]{hyperref}
\usepackage{subcaption}
\usepackage{float}
\usepackage{svg}
\usepackage{array}
\usepackage{multirow}

\title{Focal plane modal estimation of high-order wavefront drifts on coronagraphic testbeds}

\author[a]{Saikrishna Manojkumar}
\author[b,c]{Susan F. Redmond}
\author[d]{Leonid Pogorelyuk}
\author[a]{Christine L. Page}
\author[b,c]{Garreth Ruane}
\author[b,c]{Niyati Desai}
\author[a]{Kerri Cahoy}

\affil[a]{Massachusetts Institute of Technology, 77 Massachusetts Ave, Cambridge, MA 02139, USA}
\affil[b]{California Institute of Technology, 1216 E California Blvd, Pasadena, CA 91125, USA}
\affil[c]{Jet Propulsion Laboratory, California Institute of Technology, 4800 Oak Grove Dr, Pasadena, CA 91011, USA}
\affil[d]{Rensselaer Polytechnic Institute, 110 8th St, Troy, NY 12180, USA}

\authorinfo{Further author information: (Send correspondence to S.M.)\\S.M.: E-mail: saimanoj@mit.edu}

\begin{document} 
\maketitle
\begin{abstract}
 The upcoming Habitable Worlds Observatory requires precise wavefront stability over long observation periods to directly image Exo-Earths of interest, which are 10 orders of magnitude dimmer than their host star. Dark zone maintenance (DZM) has been proposed as a focal plane high-order wavefront sensing and control (HOWFSC) strategy to mitigate instabilities caused by thermal and mechanical vibrations that contribute to slow-evolving high-order wavefront drifts, which ultimately degrade contrast. Here we present a modal dark zone maintenance (MDZM) algorithm that uses an Extended Kalman Filter (EKF) to estimate the drifting wavefront modes in the basis of the deformable mirror (DM) actuators used for control. The estimator derives information across the entire focal plane from probe-pair images, addressing deficiencies of DZM, and can directly obtain a DM command, eliminating the need for Electric Field Conjugation (EFC) in this algorithm and the regularization parameter usually used for control. MDZM has been demonstrated in simulation and in hardware on the Decadal Survey Testbed (DST) at NASA JPL. These experiments have maintained a normalized intensity (NI) of $\sim3\times10^{-10}$ for over 36 hours, and have reduced the gap between the photon-derived theoretical bounds as compared to DZM. Furthermore, it provides an estimate of naturally occurring drift on the testbed, providing potential system identification capabilities. To evaluate the robustness of MDZM, a parameter scan is conducted of MDZM on DST, spanning exposure times and drift rates. Additionally, realistic observing scenarios for HWO will involve broadband focal-plane images with a 20\% bandwidth, rather than the 2.5\% narrowband images previously used in MDZM demonstrations on testbeds. Therefore, we tested the MDZM formulation in simulation using broadband sensing images with a 10\% bandwidth, using the FALCO software package with a model based on DST, and compared its performance against MDZM runs using narrowband images.
\end{abstract}

\keywords{High-contrast imaging, coronagraph, wavefront modes, deformable mirror, modal dark zone maintenance, exoplanet detection, wavefront sensing and control, deformable mirror drift, broadband sensing}

\section{INTRODUCTION}
\label{sec:intro}  
The search for life outside of Earth is one of the primary scientific goals of the upcoming Habitable Worlds Observatory (HWO), the next flagship large-aperture observatory. This mission, currently slated for launch in the 2040s, will attempt to directly image and obtain the spectra of 25 Earth-like exoplanets, looking for signs of life such as liquid water and fulfilling one of the key recommendations from the 2020 Decadal Survey on Astronomy and Astrophysics \cite{feinberg_habitable_2024, committee_for_a_decadal_survey_on_astronomy_and_astrophysics_2020_astro2020_pathways_2023}. These exoplanets, residing in the so-called ``habitable zone", tend to be 10 orders of magnitude fainter than their host star, necessitating technologies that enable the suppression of starlight to reach an imaging contrast of $10^{-10}$\cite{stark_exoearth_2019}. 

Coronagraphs have demonstrated the ability to suppress starlight for both ground and space based telescopes. They typically consist of a focal-plane mask to occult the starlight and pupil-plane masks to address diffraction effects. The in-vacuum Decadal Survey Testbed (DST) at NASA JPL has shown this coronagraph technology to be capable of achieving a mean contrast of $4\times10^{-10}$ with the aid of two deformable mirrors (DMs) for wavefront control, thus creating a region of high-contrast known as the dark zone (DZ) \cite{patterson_design_2019, seo_testbed_2019}. This demonstration reached this contrast by correcting static aberrations present in the optical system. However, environmental disturbances, such as thermal and mechanical instabilities, can create DM actuator drifts, leading to quasi-static spatially high-order wavefront perturbations \cite{mejia_prada_high-contrast_2019}. This was observed on DST, with quasi-static aberrations leading to a DZ contrast degradation rate of $1\times10^{-10}/\text{hr}$\cite{seo_testbed_2019, meeker_twin_2021}.

HWO requires long observations periods to collect a sufficient number of photons to obtain the spectra of exoplanets. Hence, it is imperative that the DZ contrast of $10^{-10}$ is maintained throughout observation. This motivates the use of high-order wavefront sensing and control (HOWFSC) techniques to maintain contrast and enable continuous observation of the science target while correcting these aberrations which create speckles in the  image~\cite{racine_speckle_1999}. For the Nancy Grace Roman Space Telescope (RST), the Coronagraph Instrument (CGI) on-board will aim to image Jupiter-like exoplanets at a contrast of $\sim10^{-8}$\cite{cady_high-order_2025}. This will be achieved by targeting the telescope at a bright reference star to dig the DZ, and then subsequently slewing over to the target star to perform science observations. As the DZ degrades due to these quasi-static aberrations, the telescope will slew intermittently to the reference star to re-dig the DZ, thus reducing the scientific observing time available for the target~\cite{kasdin_nancy_2020}. For HWO, contrast-maintenance HOWSFSC algorithms have been proposed that remove the need for periodic slewing. 

Several algorithms have been proposed to maintain contrast in the DZ, such as Linear Dark Field Control (LDFC)\cite{miller_spatial_2017} and the self-coherent camera (SCC) \cite{Mazoyer_Baudoz_Galicher_Rousset_2014}. A focal plane technique that has been proposed is Dark Zone Maintenance (DZM), an algorithm which uses probed science exposures to estimate and null the drifting electric field in the DZ using DMs\cite{pogorelyuk_dark_2019, redmond_implementation_2022}. The probe commands, applied through the DMs, provide the phase diversity needed to estimate the electric field from intensity measurements. The algorithm is then able to estimate this electric field using an Extended Kalman Filter (EKF), and feeds the estimate to an Electric Field Conjugation (EFC) controller\cite{giveon_electric_2007} to obtain the nulling DM command. DZM has been demonstrated on coronagraph testbeds, including DST\cite{redmond_dark_2022, redmond_dark_2021, redmond_advancing_2026}. This algorithm has also been tested under various disturbance and noise magnitudes to illustrate the sensitivity of this algorithm \cite{manojkumar_parameter_2024}. This analysis tested various drift rates, which were injected onto the DMs throughout the maintenance experiments as a random walk noise profile, as well as various photon conditions in the DZ. In addition, various probe, or dither, magnitudes were applied as random zero-mean Gaussian profiles on the DMs, with the standard deviation prescribed. The results underscored a fundamental limitation of DZM, relative to the theoretical photon bounds of contrast, derived from the Cramer-Rao bound\cite{ pogorelyuk_information-theoretical_2021}.

\begin{figure}[htb]
    \centering
\includegraphics[width=.8\textwidth]{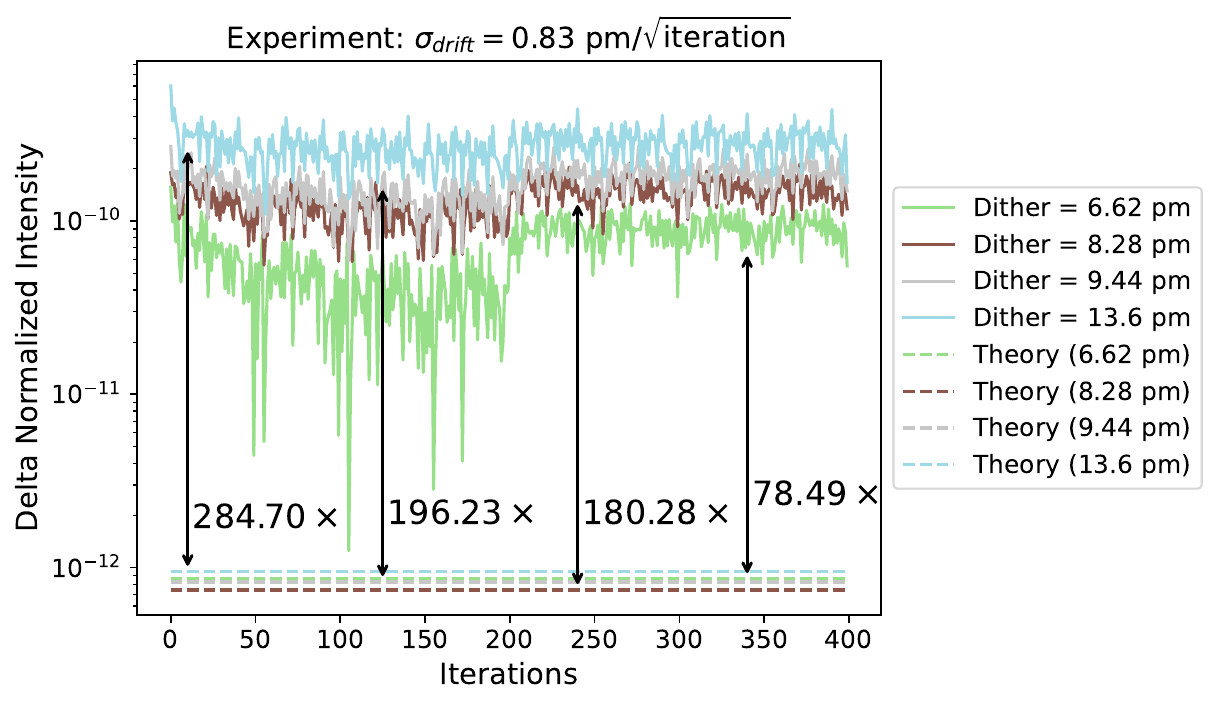}
    \caption{\label{fig:dst_dzm_theory} Iteration history of DZM runs on DST for a given drift rate, testing 4 different probe, or dither, magnitudes. The results show the delta normalized intensity (NI), the difference between the initial starting NI and achieved NI, and compare this to the theoretical bounds, highlighting the large gap present.}
\end{figure}

The DZM contrast discrepancy from the theoretical bounds primarily stems from the limitations introduced by sensing and estimating in the pixel-basis, attempting to estimate the real and imaginary components of the electric field at each pixel of the science camera. In addition, the DZM formulation, due to its block diagonal covariance structure, is not able to fully capture correlations present between pixel measurements. Hence, the large number of parameters being sensed, combined with sensitivity to photon noise, necessitates larger dither magnitudes to obtain sufficiently accurate estimates, thereby degrading contrast. This is illustrated in Figure~\ref{fig:dst_dzm_theory}, highlighting for a given rate of injected drift applied on DST, while testing various probe magnitudes, what the DZM performance provides. The results are shown as normalized intensity (NI), a common proxy of contrast. The delta NI shown, illustrating the difference between the starting NI of the DZM run and the achieved NI, is close to 2 orders of magnitude worse than the theoretical bounds across all dither magnitudes, highlighting a gap that needs to be addressed.

An alternative maintenance algorithm has been proposed to close this gap, a ``modal" dark zone maintenance (MDZM) algorithm, introduced in Pogorelyuk et al. 2022~\cite{pogorelyuk_dark_2022}. This approach was subsequently tested in simulation in Manojkumar et al. 2025~\cite{manojkumar_modal_2025} as a continuous algorithm across various drift, photon rate, and dither conditions. The algorithm was then demonstrated for the first time on hardware and in-vacuum on DST, maintaining a NI of $\sim 3\times10^{-10}$ on the testbed, overcoming injected drift and naturally occurring drift for over 36 hours, as described in Manojkumar et al. 2026~\cite{manojkumar_demonstration_2026}. These results demonstrate the potential to maintain the required 10 orders of starlight suppression with minimal degradation to the NI. However, further investigation on hardware is needed to evaluate the algorithm's robustness under varying noise levels and photon conditions, as well as its performance when using measurement images with a broader spectral bandwidth.

The remainder of this work is structured as follows: Section~\ref{sec:Modal_EKF} summarizes the Modal EKF algorithm and the previous demonstrations of MDZM on hardware; Section~\ref{sec:MDZM_param_scan} describes the various drift and exposure times which were tested on DST in the parameter scan and analyzes the results; Section~\ref{sec:broadband} discusses the broadband measurements for the algorithm and its performance on the FALCO simulator; and Section~\ref{sec:conclusion} provides conclusions and outlines future directions.

\section{MODAL EKF}\label{sec:Modal_EKF}
The MDZM algorithm differs from the traditional DZM algorithm as it attempts to estimate the drifting electric field in a latent modal basis, namely the DM actuator basis, as opposed to the pixel-basis.

\subsection{Algorithm Review}
\label{sec:algo}
The modal estimation of the electric-field uses an $r$-dimensional orthogonal basis to represent the electric field. The electric-field detected at the camera can be represented as \begin{equation}\label{eqn: Modal EKF electric field}
    \mathbf{E(u)} = \mathbf{J}\mathbf{u} + \mathbf{G}\mathbf{v} + \mathbf{E}_{\perp} \, ,
\end{equation}
where $\mathbf{J}\in \mathbb{R}^{2n \times m}$ is the Jacobian which maps $m$-dimensional DM commands, $\mathbf{u}\in \mathbb{R}^{m \times 1}$, to the $2n$-dimensional electric field (representing the real and imaginary components), noting $n$ being the number of pixels on the camera. $\mathbf{G} \in \mathbb{R}^{2n \times r}$ is the operator mapping the modal basis to the electric field, $\mathbf{v} \in \mathbb{R}^{r \times 1}$ represents the modal coefficients, and $\mathbf{E}_{\perp}\in\mathbb{R}^{2n \times 1}$ represents the rest of the wavefront that cannot be defined by the DM contributions or the modal wavefront error \cite{pogorelyuk_dark_2022}. The modal coefficients, $\mathbf{v}$, can be estimated from a pair of images with positive-negative dither commands applied, $\pm\delta\mathbf{u}_p$, analogous to pairwise-probing\cite{giveon_pair-wise_2011}. The difference of images is defined as \begin{equation}\label{eqn: pairwise_1}
    \bm{I}(\delta\mathbf{u}_p) - \bm{I}(-\delta{\mathbf{u}_p}) = 4 \mathbf{B}\,\text{diag}(\mathbf{J}\delta\mathbf{u}_p)(\mathbf{J}\mathbf{u}+\mathbf{G}\mathbf{v}+ \mathbf{E}_{\perp}) \, ,
\end{equation}
where $\mathbf{B} \in \mathbf{R}^{n \times 2n}$ is the matrix which sums intensity values across channels of the real and imaginary electric field contributions for each pixel. This matrix is defined as $\mathbf{B} = \mathbf{I}\otimes\mathbf{1}$, with $\mathbf{I}$ being the identity matrix, $\otimes$ being the Kronecker product, and $\mathbf{1}$ being a vector of ones. The drifting electric field is then simply defined as $\delta\mathbf{E} \triangleq  \mathbf{G}\mathbf{v}$.

If we explicitly define this modal basis to represent the modal coefficients as the DM basis, making the modal functions align with the DM actuators, the modal mapping matrix, $\mathbf{G}$ is equal to the Jacobian $\mathbf{J}$, allowing for Equation~\ref{eqn: pairwise_1} to be rewritten as \begin{equation}\label{eqn:modal_measurement}
    \bm{I}(\delta\mathbf{u}_p) - \bm{I}(-\delta{\mathbf{u}_p}) = 4 \mathbf{B}\,\text{diag}(\mathbf{J}\delta\mathbf{u}_p)(\mathbf{J}(\mathbf{u}+\hat{\mathbf{v}} )) \, ,
\end{equation}
with the important note that the $\mathbf{E}_{\perp}$ term corresponding to the uncontrollable electric field which is not measurably modulated by the DMs is not in the estimate, thus biasing it. With this measurement model defined, the EKF is then formulated, estimating the state vector of the modal coefficients, $\hat{\mathbf{x}} = \hat{\mathbf{v}} \in \mathbb{R}^{m \times 1}$, with an initial covariance of $\mathbf{P}_0\in \mathbb{R}^{m \times m}$. Like DZM, each iteration, $k$, assumes static dynamics such that $\hat{\mathbf{x}}_{k+1|k} = \hat{\mathbf{x}}_k$. The covariance is propagated as $\mathbf{P}_{k+1|k} = \mathbf{P}_k + \mathbf{Q}_v$ with the process noise covariance matrix, $\mathbf{Q}_v$, being defined as \begin{equation}\label{eqn: process_noise}
    \mathbf{Q}_v = \sigma_{drift}^2\mathbf{I} \, ,
\end{equation}
where $\sigma_{drift}$ is the standard deviation of the system drift rate, accounting for injected DM drift and naturally occurring testbed and DM drift effects. The measurement model is used to update the state and covariances with the probed images, $\mathbf{y} (\pm\delta\mathbf{u}_p)$. The observation matrix $\mathbf{H}_k \in \mathbb{R}^{n \times m}$ maps the measurements to the state as \begin{equation}
    \mathbf{H}_k = \frac{\partial(\mathbf{y}(\delta\mathbf{u}_p) - \mathbf{y}(-\delta\mathbf{u}_p))}{\partial \mathbf{v}_k} = 4\mathbf{B}\,\text{diag}(\mathbf{J}\delta\mathbf{u}_p)\mathbf{J} \, .
\end{equation}
The measurement noise covariance is also carefully defined as \begin{equation}\label{eqn:mdzm_r}
    \mathbf{R}_k = \text{diag}\left(\mathbf{y}(\delta\mathbf{u}_p) + \mathbf{y}(-\delta\mathbf{u}_p) + 2(\mu_I t_{exp})^2 + 2\sigma_{rn}^2\right) \, ,
\end{equation}
a function of the measurement images, the dark current $\mu_I$, the exposure time $t_{exp}$, and the read noise $\sigma_{rn}$.

\begin{figure}[htb]
    \centering
    \includegraphics[width=\textwidth]{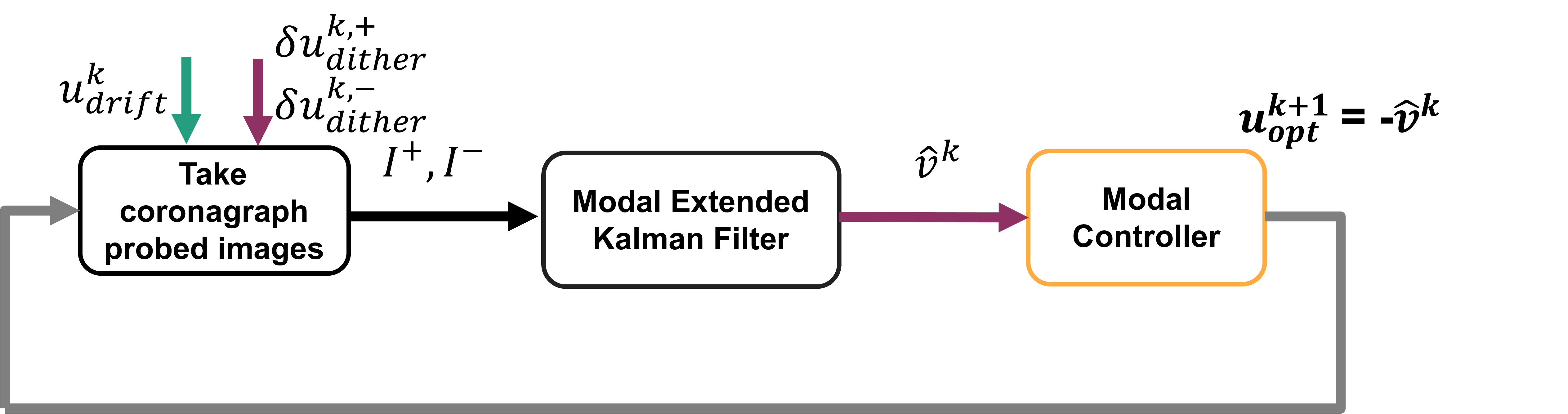}
    \caption{\label{fig:mdzm_diagram} Diagram illustrating the control loop process of the MDZM algorithm.}
\end{figure}

The final measurement, $\mathbf{y} = \mathbf{y}(\delta\mathbf{u}_p) - \mathbf{y}(-\delta\mathbf{u}_p)$, is compared to the predicted measurement $\hat{\mathbf{y}} = \hat{\mathbf{y}}(\delta\mathbf{u}_p) - \hat{\mathbf{y}}(-\delta\mathbf{u}_p)$, which can be defined from Equation~\ref{eqn:modal_measurement}, as \begin{equation}
\mathbf{\hat{y}}(\pm \delta\mathbf{u}_p) = \Re\left(\mathbf{J}(\mathbf{u} + \mathbf{\hat{v}} \pm \delta\mathbf{u}_p)\right) \circ \Re\left(\mathbf{J}(\mathbf{u} + \mathbf{\hat{v}} \pm \delta\mathbf{u}_p)\right) + \Im\left(\mathbf{J}(\mathbf{u} + \mathbf{\hat{v}} \pm \delta\mathbf{u}_p)\right) \circ \Im\left(\mathbf{J}(\mathbf{u} + \mathbf{\hat{v}} \pm \delta\mathbf{u}_p)\right) \, .
\end{equation}
The calculated Kalman gain, \begin{equation}\label{eqn:kalman_gain}
    \mathbf{K}_k = \mathbf{P}_{k+1|k}\mathbf{H}_k^T(\mathbf{H}_k\mathbf{P}_{k+1|k}\mathbf{H}_k^T + \mathbf{R}_k)^{-1} \, ,
    \end{equation}
updates the state vector as \begin{equation}
    \hat{\mathbf{x}}_{k+1} = \hat{\mathbf{x}}_{k+1|k} + \mathbf{K}_k(\mathbf{y} - \hat{\mathbf{y}}),
\end{equation} and covariance as \begin{equation}
    \mathbf{P}_{k+1} = (\mathbf{I} - \mathbf{K}_k\mathbf{H}_k)\mathbf{P}_{k+1|k}.
\end{equation}
With the drifting electric field directly mapped to the DM basis, the optimal command to maintain contrast can simply be found as the negation of the estimated coefficients, \begin{equation}\label{eqn: dm_command}
    \mathbf{u} = -\hat{\mathbf{x}} \, ,
\end{equation}
removing the need for EFC. A full summary of the MDZM process is shown in the diagram in Figure~\ref{fig:mdzm_diagram}.

\subsection{Demonstrations}\label{sec:demonstrations}
\begin{figure}[htb]
    \centering
\includegraphics[width=.6\textwidth]{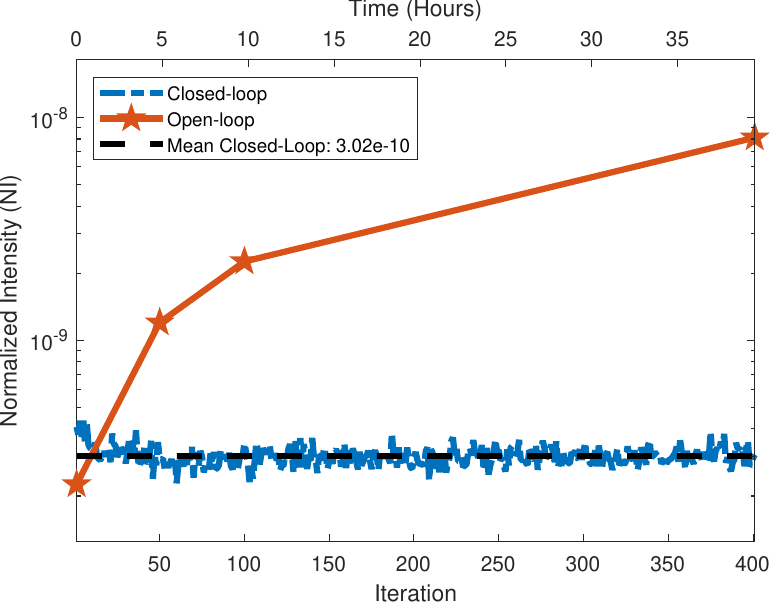}
    \caption{\label{fig:MDZM_DST_old} Iteration history of a MDZM run on DST for a given injected drift magnitude of $1.425\times10^{-3} ~\text{V}/\sqrt{\text{iter}}$ and dither magnitude of $2.8\times10^{-3} ~\text{V}/\text{iter}$\cite{manojkumar_demonstration_2026}.}
\end{figure}

\begin{figure}[htb]
    \centering
\includegraphics[width=.8\textwidth]{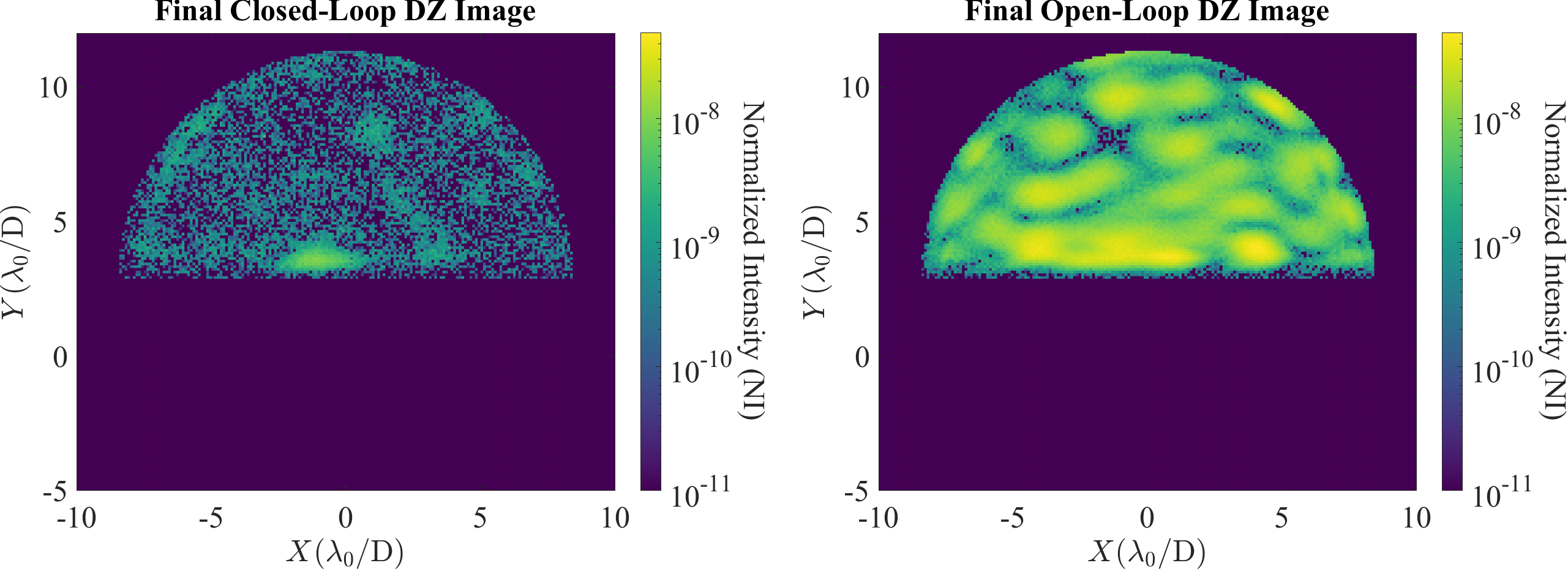}
\caption{\label{fig:contrast_comparison_44} Final closed-loop and open-loop DZ images for a MDZM run on DST with a given injected drift magnitude of $1.425\times10^{-3} ~\text{V}/\sqrt{\text{iter}}$ and dither magnitude of $2.8\times10^{-3} ~\text{V}/\text{iter}$\cite{manojkumar_demonstration_2026}.}
\end{figure}

The MDZM algorithm was previously tested in simulation, but was first tested on hardware on DST, in September 2025, as described in Manojkumar et al. 2026~\cite{manojkumar_demonstration_2026}. Three different MDZM runs were performed, each spanning over 36 hours with 120 s exposures and maintaining a NI as good as $3\times10^{-10}$. An illustrative run is shown in Figure~\ref{fig:MDZM_DST_old}, comparing the closed-loop NI with control applied to the open-loop NI, illustrating the degradation in the DZ that was suppressed by the algorithm, with this case suppressing over 1.5 orders of magnitude of degradation. 
This successful performance was further illustrated by analyzing the open-loop and closed-loop DZ images, as shown in Figure~\ref{fig:contrast_comparison_44}, indicating the successful suppression of speckles.

\begin{figure}[htb]
    \centering
    \begin{subfigure}[b]{0.45\textwidth}
        \centering
        \includegraphics[width=\textwidth]{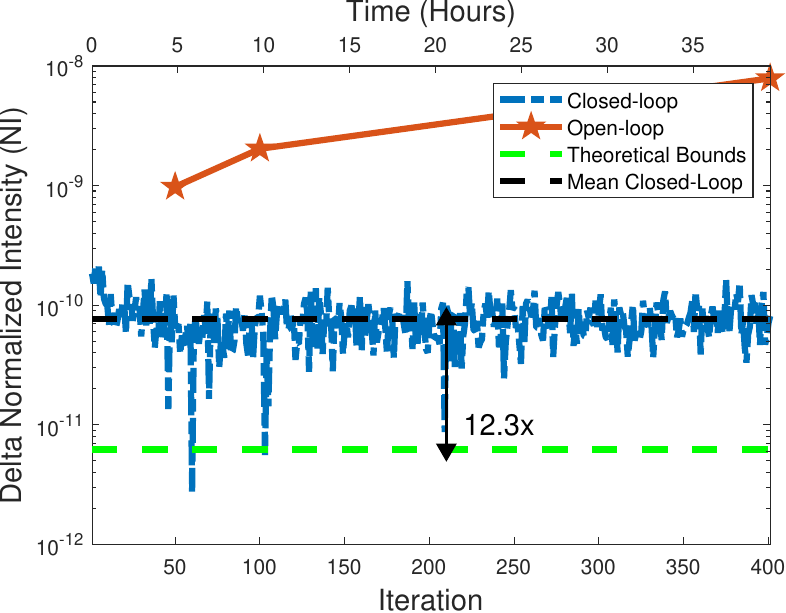}
    \end{subfigure}
    \hspace{0.05\textwidth}
    \begin{subfigure}[b]{0.45\textwidth}
        \centering
        \includegraphics[width=\textwidth]{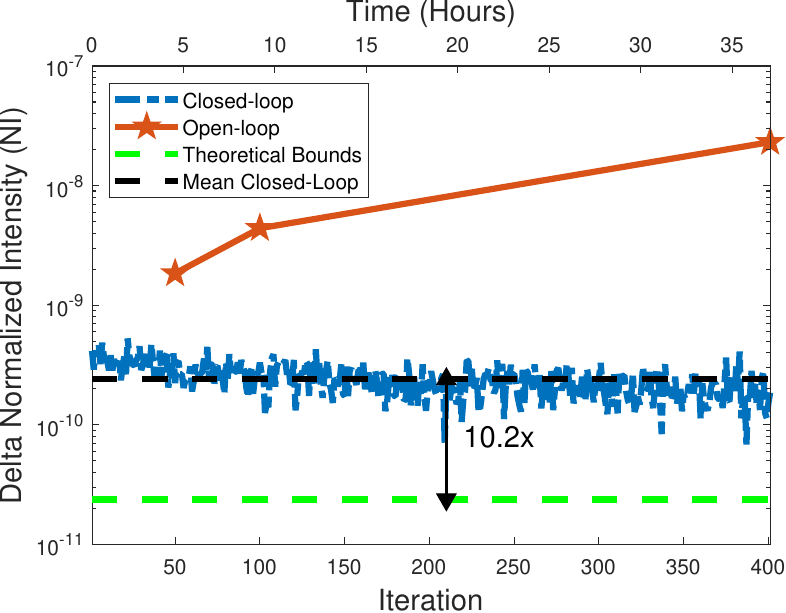}
    \end{subfigure}

    \begin{subfigure}[b]{0.45\textwidth}
        \centering
        \includegraphics[width=\textwidth]{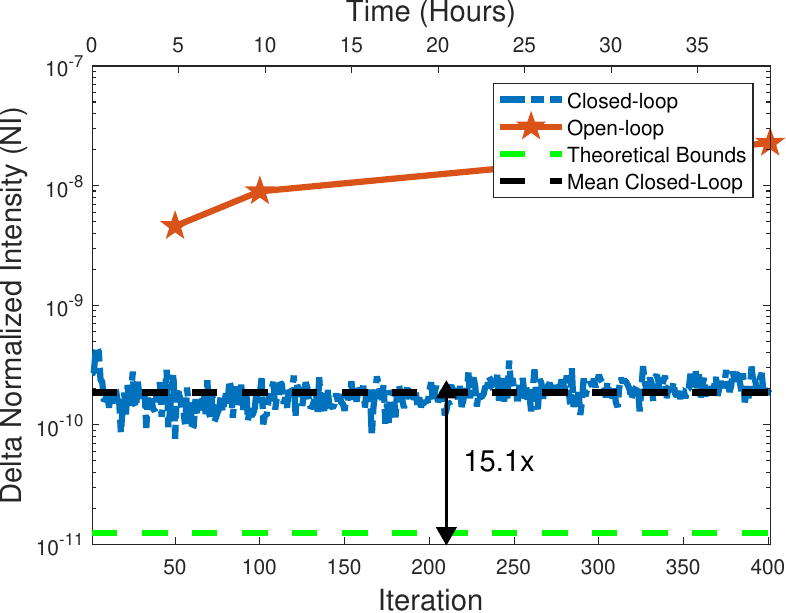}
    \end{subfigure}
    \hspace{0.05\textwidth}

    \caption{\label{fig:modal_theory} The comparison of the three MDZM experiments analyzed in Manojkumar et al. 2026~\cite{manojkumar_demonstration_2026} to the theoretical bounds. The  delta NI is shown across each iteration, with the factor difference between the mean closed-loop NI and the theoretical bounds labeled for each run.}
\end{figure}

After performing the runs on DST, it was important to compare the MDZM performance to the theoretical bounds. This was achieved by comparing the achieved delta NI, indicating the difference from the initial open-loop NI to the achieved NI, similar to Figure~\ref{fig:dst_dzm_theory}. The comparison to theory is shown in Figure~\ref{fig:modal_theory}. Each plot clearly shows that the MDZM algorithm is able to close this gap with the theoretical bounds as compared to DZM, with the factor difference between theory reducing to one order of magnitude from two. There also remains further room for improvement, as the dither magnitudes which were applied were not optimized to each best possible performance. Rather, a conservatively large dither was applied to ensure stability, while also noting hardware limitations of the DM electronics on the testbed allowing a minimum dither magnitude of $\sigma_{dither}=2.8\times10^{-3}~\text{V}/\text{iter}$.

\begin{figure}[htb]
    \centering
    \begin{subfigure}[b]{0.45\textwidth}
        \centering
        \includegraphics[width=\textwidth]{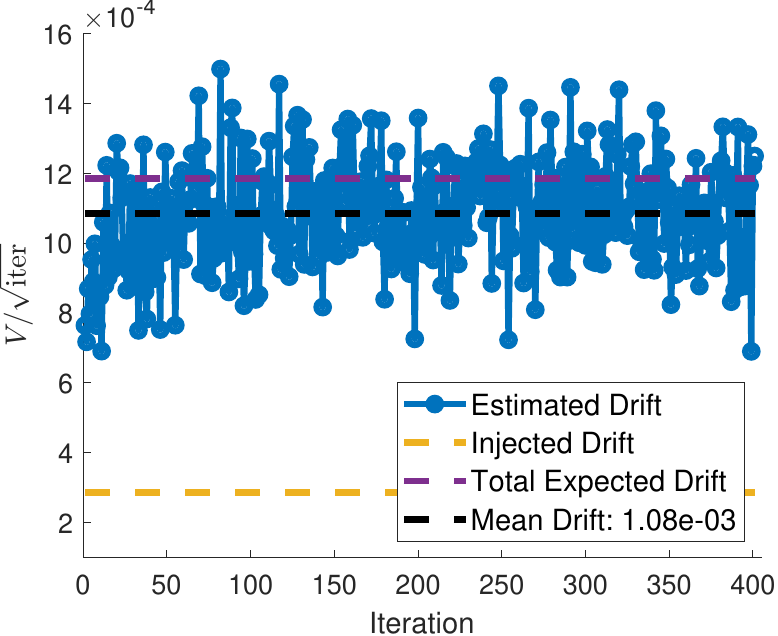}
    \end{subfigure}
    \hspace{0.05\textwidth}
    \begin{subfigure}[b]{0.45\textwidth}
        \centering
        \includegraphics[width=\textwidth]{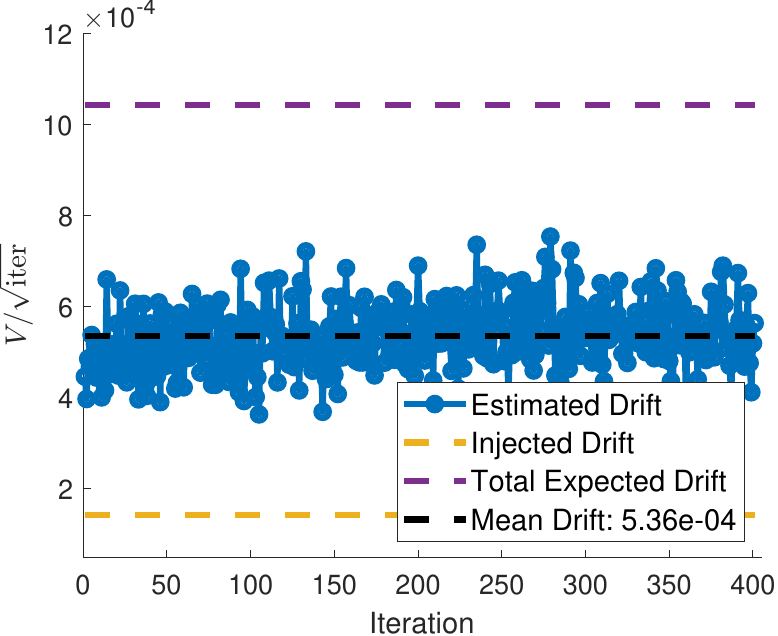}
    \end{subfigure}

    \begin{subfigure}[b]{0.45\textwidth}
        \centering
        \includegraphics[width=\textwidth]{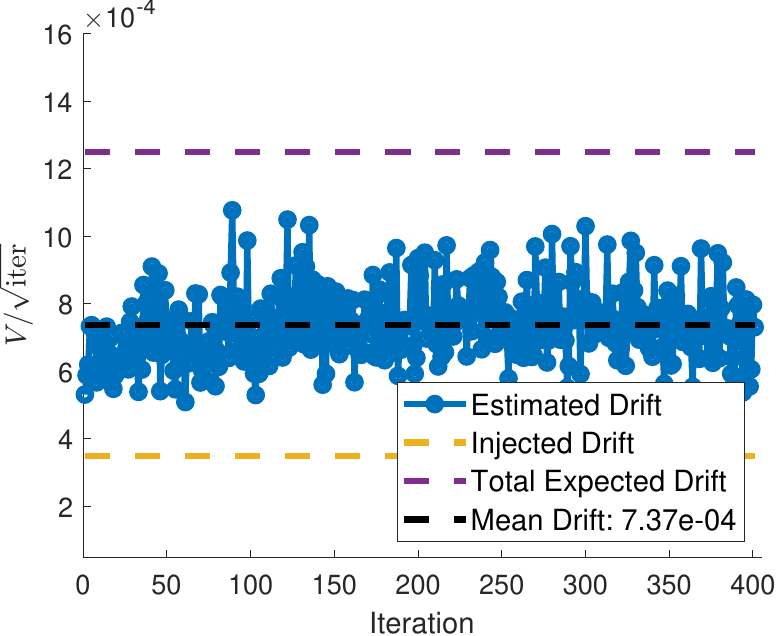}
    \end{subfigure}
    \hspace{0.05\textwidth}

    \caption{\label{fig:modal_drift_estimate} The estimated drift rates from the MDZM state vector across all experiments conducted in Manojkumar et al. 2026~\cite{manojkumar_demonstration_2026}, compared to the true injected drift and total expected drift rates.}
\end{figure}

Another interesting analysis which was conducted on the results from the runs presented in Manojkumar et al. 2026~\cite{manojkumar_demonstration_2026} was the estimated drift rate the EKF produced. In addition to the injected drift magnitude, naturally occurring drift present on the testbed, accumulating over the long experiment times, had a large impact on contrast. This naturally injected drift needed to be accounted for, and hence the process-noise covariance matrix defined in Equation~\ref{eqn: process_noise} had the drift magnitude defined as \begin{equation}
    \sigma_{drift} = \sigma_{drift,i} + \sigma_{drift,n} \, ,
\end{equation}
where $\sigma_{drift,i}$ is the injected drift and $\sigma_{drift,n}$ is the naturally occurring drift rate. These natural drift rates were set as apriori values of $9\times10^{-4}~\text{V}/\sqrt{\text{iter}}$, in the absence of true knowledge of the naturally occurring drift. Despite not knowing this, each run was able to maintain stable contrast, though it is possible that some runs suffered from having larger naturally occurring drift than others. One possible source of information for this true natural drift rate is the estimate from the MDZM, which corresponds to the drifting modes of the DMs, and hence the subsequent corrections. The standard deviation of this estimate across each of the three experiments, and the subsequent mean estimated drift for each iteration is compared to the injected drift and total expected drift, which accounts for the natural drift, as shown in Figure~\ref{fig:modal_drift_estimate}. The results show for each experiment the estimated drift rate is larger than the injected, indicating the presence of naturally occurring drift. This estimate, however, is less than the total expected drift, with the estimated natural drift varying across each experiment, though remaining quite stable across the run. This warrants further investigation to properly characterize the naturally occurring drift for the possible use of MDZM as a form of system identification for coronagraphic systems.

\section{MDZM Parameter Scan}
\label{sec:MDZM_param_scan}
\begin{figure}[htb]
    \centering
    \begin{subfigure}[b]{0.7\textwidth}
        \centering
        \includegraphics[width=\textwidth]{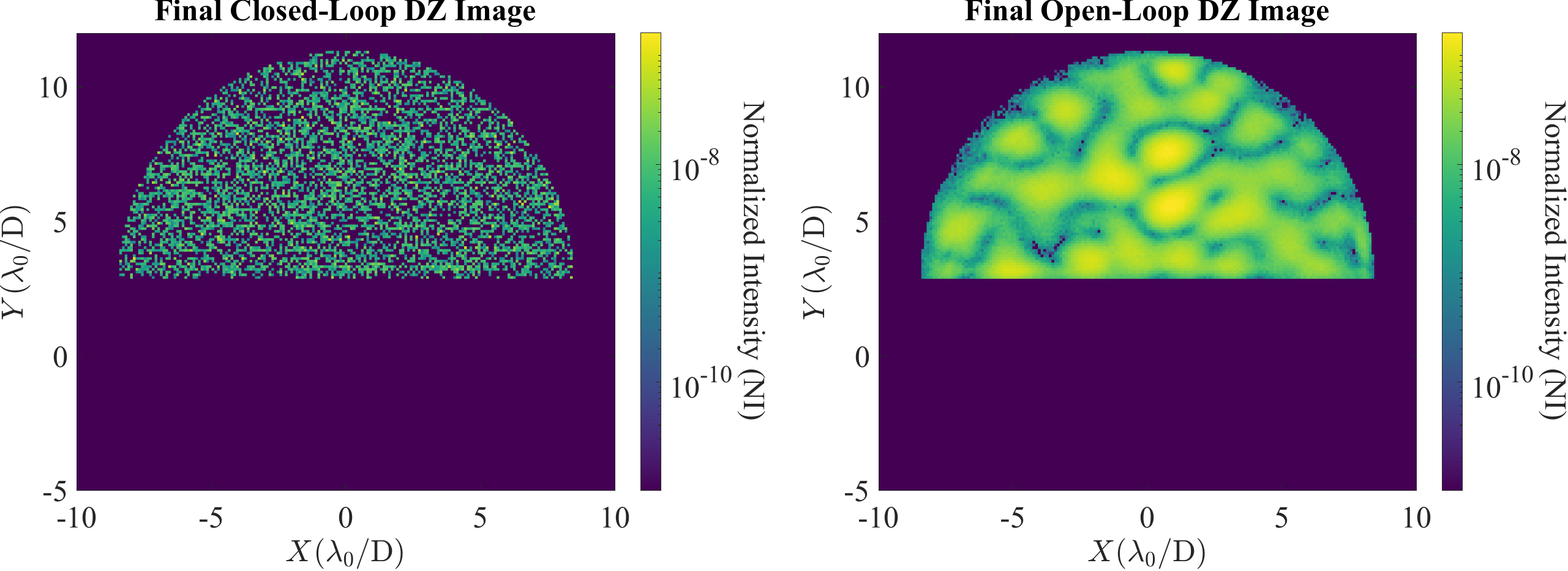}
    \end{subfigure}
    \hspace{0.05\textwidth}
    \begin{subfigure}[b]{0.7\textwidth}
        \centering
        \includegraphics[width=\textwidth]{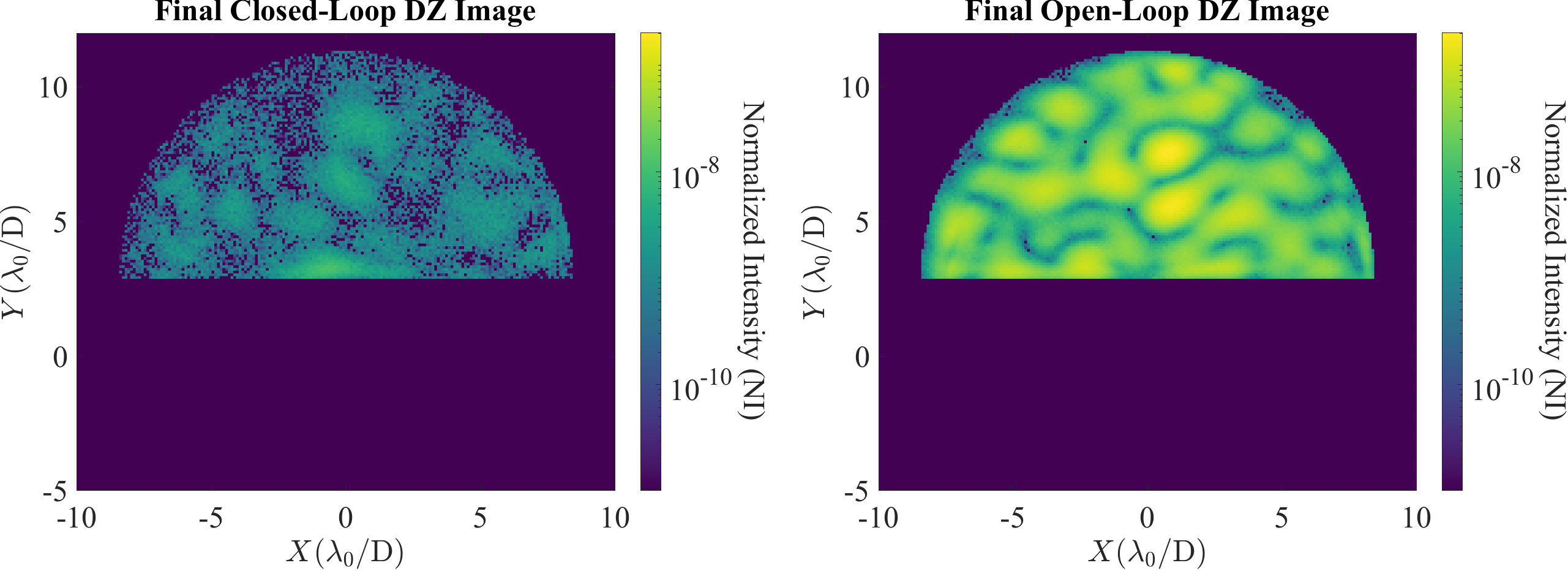}
    \end{subfigure}

    \caption{\label{fig:exposure_comparison} A comparison of the final closed-loop and open-loop DZ images for a representative 1.5 s exposure case on the top and a 90s exposure case on the bottom.}
\end{figure}

Building upon the work performed in Section~\ref{sec:demonstrations}, a parameter scan of MDZM was performed on DST, analyzing the performance across three different drift rates, as well as five different exposure times. The exposure time variable varied the number of photons available in the DZ, and hence served as a proxy for the signal-to-noise ratio (SNR), varying the quality of information the EKF had access to at each iteration. A summary of the conditions tested on DST, as well as the mean normalized intensity for each run (reported as $\times10^{-10}$) is shown in Table~\ref{tab:exposure_drift_performance}.
\begin{table}[H]
\centering
\caption{Mean normalized intensity (NI) $\times 10^{-10}$ across varying exposure times and injected drift rates.}
\label{tab:exposure_drift_performance}
\begin{tabular}{|c|*{3}{>{\centering\arraybackslash}m{2cm}|}}
\hline
\multirow{2}{*}{\textbf{Exposure Time (s)}} & \multicolumn{3}{c|}{\textbf{Injected Drift Rate ($10^{-4}$ V/$\sqrt{\text{iter}}$)}} \\
\cline{2-4}
& 3.5 & 5.25 & 15 \\
\hline
1.5 & 8.93 & 9.67 & 10.7 \\
3   & 6.19 & 6.36 & 7.37 \\
15  & 4.63 & 4.62 & 5.32 \\
45  & 5.35 & 5.18 & 5.78 \\
90  & 8.16 & 6.98 & 7.11 \\
\hline
\end{tabular}
\end{table}
Furthermore, an illustration of the impact the varying exposure times had on the quality of images taken on DST for the EKF is shown in Figure~\ref{fig:exposure_comparison}, comparing a 1.5s exposure to a 90 s exposure. The final-closed DZ image for the 1.5s exposure is shown to be extremely noisy and granular, indicating a lack of sufficient information, and hence, possibly corrupted information being fed to the EKF, as compared to the much cleaner 90s closed-loop DZ image.

\begin{figure}[htb]
    \centering
    \begin{subfigure}[b]{0.48\textwidth}
        \centering
        \includegraphics[width=\linewidth, trim={4cm 8cm 4cm 8cm}, clip]{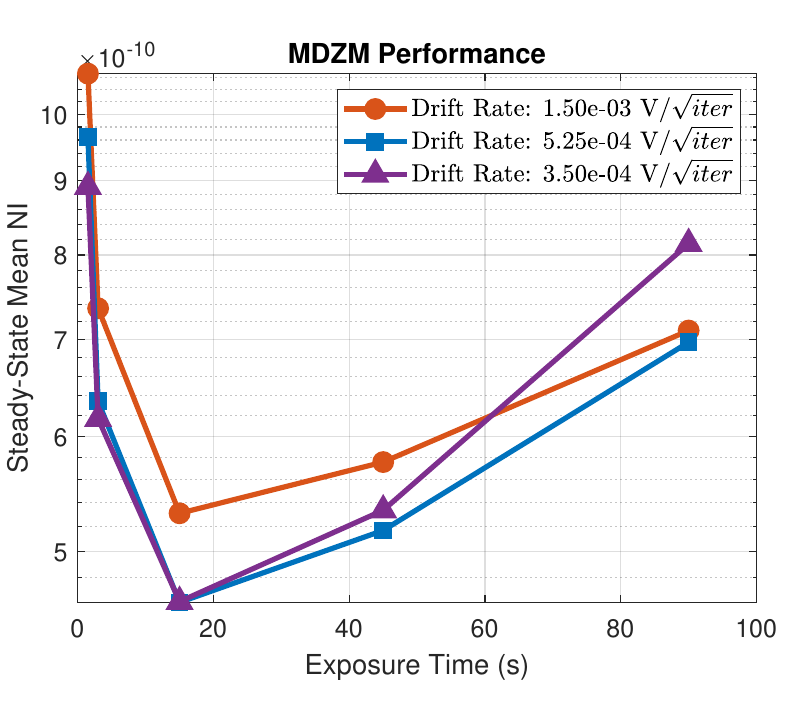}
    \end{subfigure}
    \hfill
    \begin{subfigure}[b]{0.48\textwidth}
        \centering
        \includegraphics[width=\linewidth, trim={4cm 8cm 4cm 8cm}, clip]{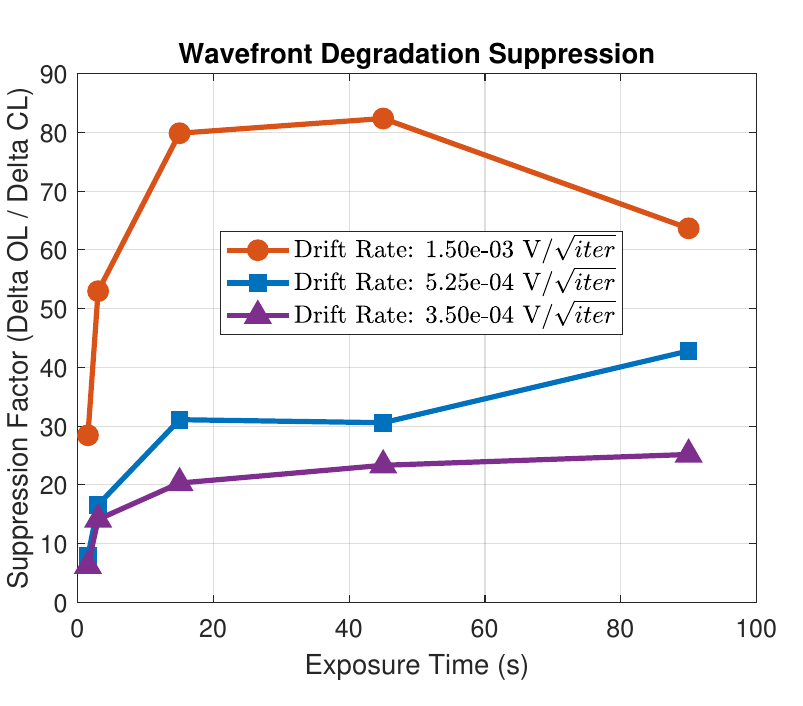}
    \end{subfigure}

    \caption{\label{fig:mdzm_paramscan} Parameter scan results for the MDZM experiments, highlighting performance across various different exposure times and drift rates. The left plot illustrates the steady-state NI achieved across each experiment, while the right plot indicates the NI degradation suppression factor, the ratio between the change in open-loop NI compared to closed-loop NI.}
\end{figure}

The mean steady-state NI achieved across each experiment as well as the NI degradation suppression factor, which is the change in the open-loop NI divided by the closed-loop NI across the experiments are shown in Figure~\ref{fig:mdzm_paramscan}. While the steady-state NI provides the actual achieved NI, the suppression factor is slightly more informative as each experiment did not start with a DZ of the same NI. The degradation suppression factor results demonstrate the improvement in performance achieved as a function of exposure time across all of the drift conditions, highlighting the importance of having sufficient photon information for the EKF to achieve an accurate estimate. The results also show that the higher drift rates are able to achieve high wavefront degradation suppression, illustrating that MDZM is able to operate even under high drift conditions, suppressing close to two order of magnitude of NI degradation. A point to note is the reduction in performance for the high drift case with a long exposure time of 90s, achieving worse performance than the 45s case. This may be due to the long exposure times introducing larger natural drift between iterations, the effect of which compounds with the large injected drift to provide slightly worse performance. However, this will need to be analyzed in the future for more definitive conclusions.

\section{Broadband MDZM}
\label{sec:broadband}
All previous demonstrations of MDZM, including the results provided in Sections~\ref{sec:demonstrations} and~\ref{sec:MDZM_param_scan}, used images with a narrow spectral bandwidth, specifically 2.5\% centered at 640 nm, to update the EKF. However, future space observatories will be imaging with a much larger spectral bandwidth. For example, the RST CGI band 1 will have a 10\% total bandwidth, with 3 subbands~\cite{cady_high-order_2025}. Furthermore, HWO will require the maintenance of a $\sim1\times10^{-10}$ contrast across a 20\% bandwidth~\cite{10.1117/12.3065725}. Hence, it is imperative that the performance of MDZM is analyzed using broadband images, rather than the narrowband. 

\begin{figure}[htb]
    \centering
    \begin{subfigure}[b]{0.48\textwidth}
        \centering
        \includegraphics[width=\textwidth]{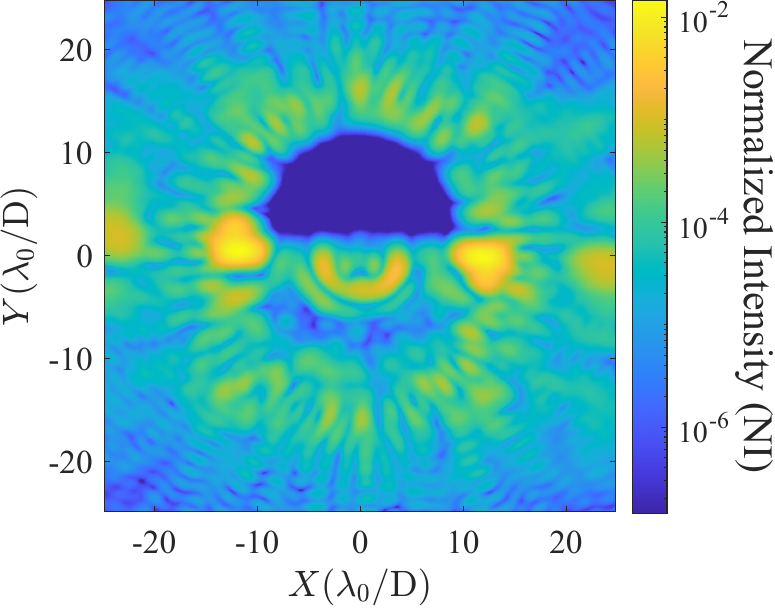}
    \end{subfigure}
    \begin{subfigure}[b]{0.48\textwidth}
        \centering
        \includegraphics[width=\textwidth]{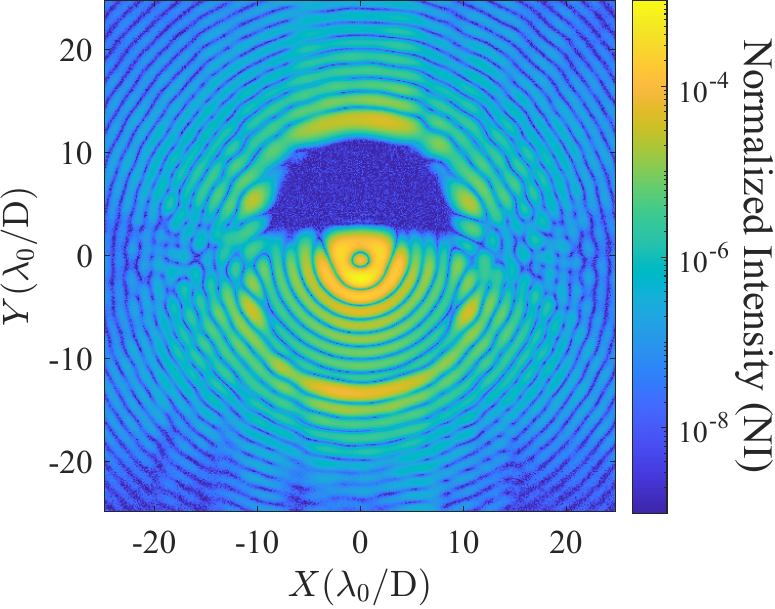}
    \end{subfigure}
    \caption{\label{fig:broadband_compare} Images taken for a 10\% (left) and 2.5\% (right) band on DST, simulated using the FALCO package, illustrating the effect of the full chromatic aberrations on the larger band images.}
    \end{figure}

The analysis of broadband MDZM was performed in simulation using the FALCO software package, simulating the performance on DST~\cite{riggs_fast_2018}. The broadband images, taken with a full 10\% band are compared to the narrowband 2.5\% band images in Figure~\ref{fig:broadband_compare}. The broadband images are shown to be brighter and to have a greater amount of blurring due to the chromatic aberrations across the band, whereas the 2.5\% band is much clearer and sharper, showing the coronagraph mask's diffraction pattern.

\begin{figure}[htb]
    \centering
    \begin{subfigure}[b]{0.48\textwidth}
        \centering
        \includegraphics[width=\textwidth]{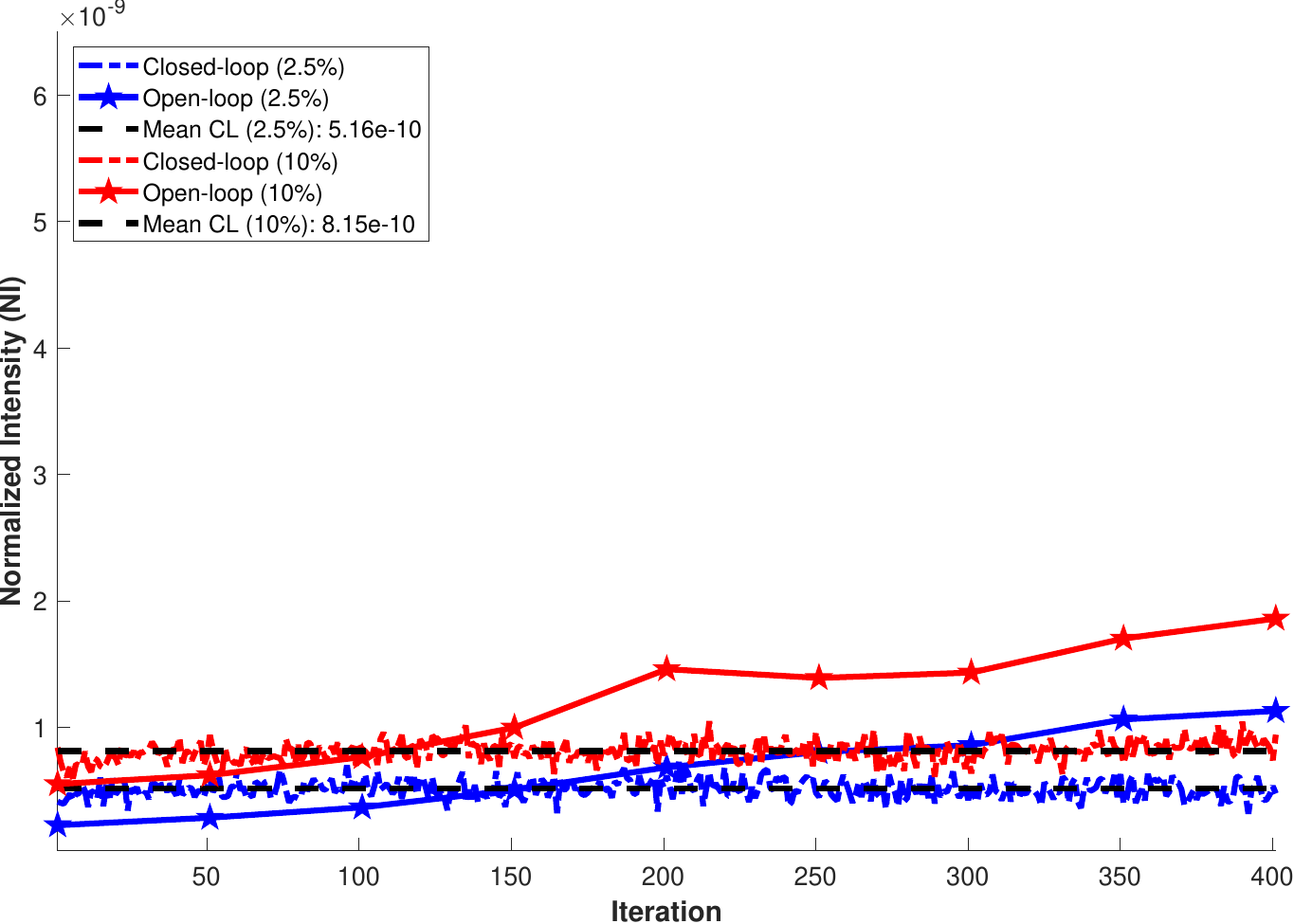}
    \end{subfigure}
    \begin{subfigure}[b]{0.48\textwidth}
        \centering
        \includegraphics[width=\textwidth]{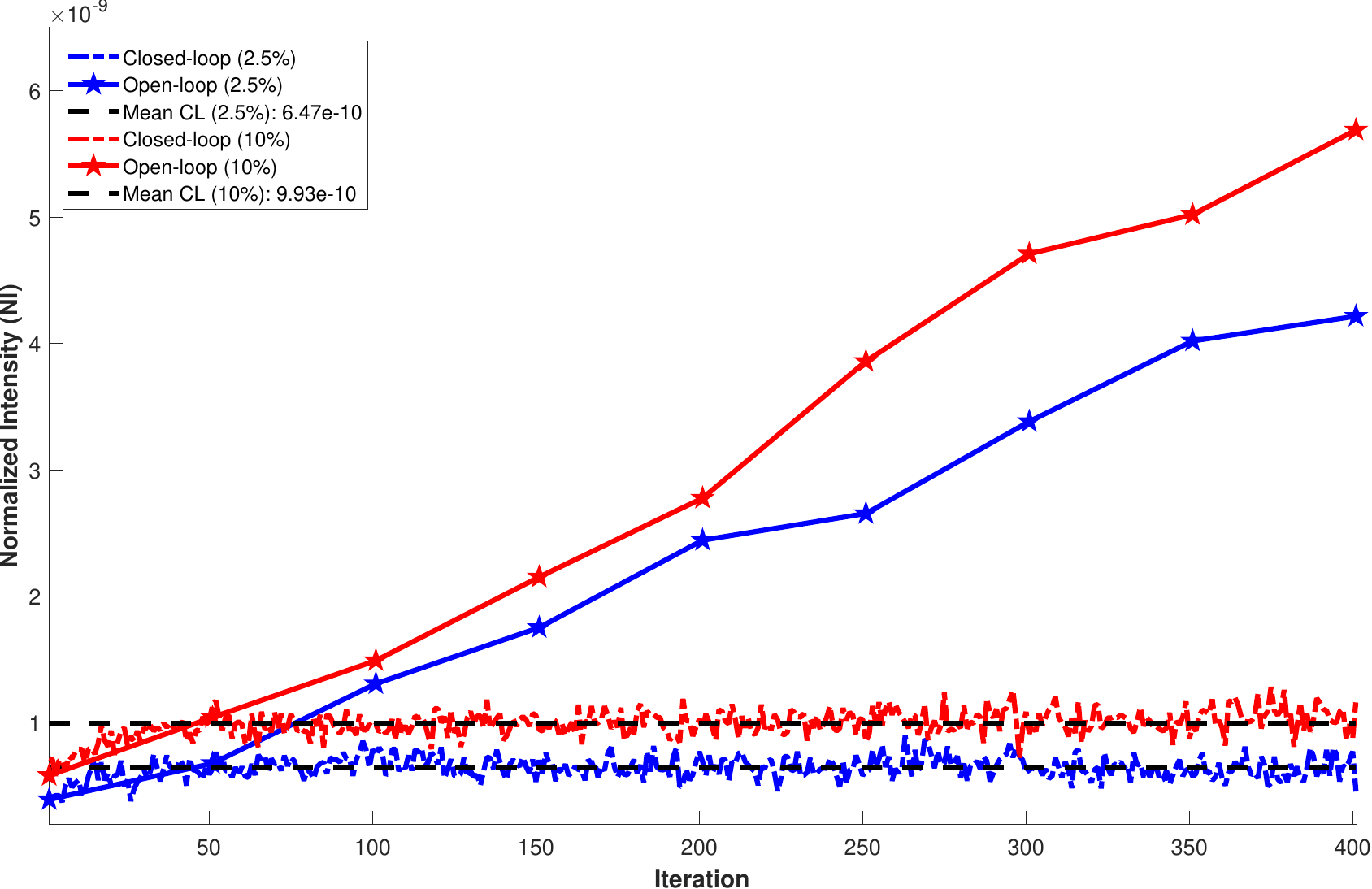}
    \end{subfigure}
    \caption{\label{fig:broadband_results} Iteration history of broadband and narrowband MDZM experiments, highlighting the open-loop and closed-loop NI achieved, for a injected drift rate of $\sigma_{drift,i} = 3.5\times10^{-4}~\text{V}/\sqrt{\text{iter}}$ (left) and $\sigma_{drift,i} = 7\times10^{-4}~\text{V}/\sqrt{\text{iter}}$ (right). A constant dither magnitude of $\sigma_{dither} = 2.8\times10^{-3}~\text{V}/\text{iter}$ was applied for each experiment.}
\end{figure}

Finally, we can compare the effect these broadband images have on the overall performance of MDZM as compared to the narrowband. The broadband images were fed as measurements to MDZM, while using the same Jacobian as the narrowband images, built around a wavelength of 640 nm. Results comparing the two bandwidth of images, for two different injected drift rates, $\sigma_{drift,i} = 3.5\times10^{-4}~\text{V}/\sqrt{\text{iter}}$ and $\sigma_{drift,i} = 7\times10^{-4}~\text{V}/\sqrt{\text{iter}}$, while using the same dither magnitude of $\sigma_{dither} = 2.8\times10^{-3}~\text{V}/\text{iter}$ are shown in Figure~\ref{fig:broadband_results}, highlighting the open-loop and closed-loop NI. The results clearly illustrate the effect the higher drift rate has on both experiments, with the higher drift rates leading to greater open-loop NI degradation. However, the closed-loop NI provided nearly identical performance, showing the effectiveness of MDZM in maintaining contrast in-spite of drift while mainly being sensitive to dither magnitude. 

When comparing the broadband to narrowband performance, broadband consistently performs worse, across both drift rates and for open-loop and closed-loop. The open-loop degradation illustrates that summed across a larger band, there are more aberrations present than what was originally captured across the narrowband image. On the other hand, the inability for MDZM to maintain a $\sim10^{-10}$ NI across the broadband experiments indicates that MDZM is highly sensitive to the chromatic nature of higher-order aberrations present, and the Jacobian constructed around 640 nm is simply not sufficient to estimate all of the aberrations. Hence, future experiments will need to refine the estimation structure to deal with broadband images.

\section{CONCLUSION and FUTURE WORK}\label{sec:conclusion}
The MDZM algorithm was further developed in this work, building upon the first simulation demonstration of this algorithm in Manojkumar et al. 2025~\cite{manojkumar_modal_2025}, and the first hardware demonstration in-vacuum on DST at NASA JPL in Manojkumar et al. 2026~\cite{manojkumar_demonstration_2026}. The results of MDZM on DST show an improvement in performance compared to traditional DZM, specifically illustrating a smaller gap between the Cramer-Rao theoretical bounds derived from available photon information, and the NI achieved, moving from an average of 2 orders of magnitude gap to 1. Furthermore, the ability of the MDZM algorithm to provide a source of information of the natural testbed drift using the estimate derived was also illustrated, motivating further investigation of using MDZM as a system identification tool. Finally, the analysis of MDZM under varying drift rates and exposure times, as well as using broadband images, highlighted the operating limitations of the current configuration of MDZM, and areas that require further refinement. 

Future work will look to test the broadband images with MDZM on hardware, and refining the MDZM formulation to allow this algorithm to take in broadband images, potentially estimating subbands within the 10\% band. Furthermore, while the experiments presented in this work used a constant dither magnitude and a Gaussian shape, it is possible to dynamically estimate the optimal dither magnitude used in the MDZM experiment, as highlighted in Manojkumar et al. 2025~\cite{manojkumar_modal_2025}. Hence, future work will look at applying this ``optimal dither" algorithm on hardware, and potentially testing various dither shapes, with the known lower hardware limitation for dither magnitude of $2.8\times 10^{-3} \text{V}/\text{iter}$ for DST. Additionally, it has also been noted in Pogorelyuk et al. 2022~\cite{pogorelyuk_dark_2022} that the number of modes estimated by MDZM is a variable that can be optimized to lower computational burden and better condition the estimation schema, requiring further investigation. The process noise covariance in each experiment was statically prescribed apriori and not updated during the experiment. Using the estimation of natural and injected drift MDZM provides, and illustrated in this work, it is possible to dynamically update the covariance during the experiment to provide better statistical conditioning. Finally, to provide a conclusive roadmap for implementation on HWO, MDZM performance needs to be projected to analyze feasibility.

\appendix    

\acknowledgments 
This work was supported by the National Aeronautics and Space Administration under Grant NNH21ZDA001N issued through the Strategic Astrophysics Technology/Technology Demonstration for Exo-planet Missions Program (SAT-TDEM; PI: K. Cahoy). The research was carried out in part through the Jet Propulsion Laboratory, California Institute of Technology, under a contract with the National Aeronautics and Space Administration (80NM0018D0004). 
\bibliography{report} 
\bibliographystyle{spiebib} 

\end{document}